# Simple method for determining binding energies of fullerene negative ions

Z. Felfli and A.Z. Msezane
Department of Physics and Center for Theoretical Studies
of Physical Systems,
Clark Atlanta University, Atlanta, US

**Abstract**
A robust potential wherein is embedded the crucial core-polarization interaction is used in the Regge-pole methodology to calculate low-energy electron elastic scattering total cross section (TCS) for the $C_{60}$ fullerene in the electron impact energy range $0.02 \leq E \leq 10.0$ eV. The energy position of the characteristic dramatically sharp resonance appearing at the second Ramsauer–Townsend (R-T) minimum of the TCS representing stable $C_{60}^-$ fullerene negative ion formation agrees excellently with the measured electron affinity (EA) of $C_{60}$ [Huang *et al* 2014 *J. Chem. Phys.* **140** 224315]. The benchmarked potential and the Regge-pole method are then used to calculate electron elastic scattering TCSs for selected fullerenes, from $C_{54}$ through $C_{240}$. The TCSs are found to be characterized generally by R-T minima, shape resonances (SRs) and dramatically sharp resonances representing long-lived ground state fullerene negative ion formation. For the TCSs of $C_{70}$, $C_{76}$, $C_{78}$, and $C_{84}$ the agreement between the energy positions of the very sharp resonances, corresponding to the binding energies (BEs) of the resultant fullerene negative ions, and the measured EAs is outstanding. Additionally, we extract the BEs of the resultant fullerene negative ions from our calculated TCSs of the $C_{86}$, $C_{90}$ and $C_{92}$ fullerenes with estimated EAs $\geq 3.0$ eV by the experiment [Boltalina *et al*, 1993 Rapid Commun. Mass Spectrom. **7** 1009] as well as of other fullerenes, including $C_{180}$ and $C_{240}$. Most of the TCSs presented in this paper are the first and only; our novel approach is general and should be applicable to other fullerenes as well and complex heavy atoms, such as the lanthanide atoms. We conclude with a remark on the catalytic properties of the fullerenes through their negative ions.

## 1. Introduction

Progress toward a fundamental theoretical understanding of the mechanism underlying low-energy electron scattering from heavy and complex atoms, including fullerenes, leading to stable negative ion formation has been very slow. The presence of many intricate and diverse electron configurations that characterize low-energy electron interactions in these systems leads to computational complexity that render very difficult, if not impossible to obtain reliable electron affinities (EAs) for complex systems. Electron affinities calculated using structure-based theoretical methods tend to be riddled with uncertainties. Accurate and reliable atomic and molecular affinities are essential for understanding chemical reactions involving negative ions [1], whose importance and utility in terrestrial and stellar atmospheres as well as in device fabrication are well-documented [2-6]. Due to their extraordinary practical properties, fullerenes are crucial to nanotechnology and industrial research. Their extensive and important applications in science and technology are reported in for example [7] and references therein. These include solar cells, optical applications, strengthening and hardening of metals, fullerenes as molecular wires and water-soluble fullerene compounds in medicine.

Low-energy electron scattering cross sections from complex atoms, characterized generally by Ramsauer–Townsend (R-T) minima, shape resonances (SRs) and dramatically sharp resonances representing stable negative ion formation [8], could yield deep insight into quantum dynamics through reliable calculations of the EAs [9]. At the R-T minimum the electrons pass through as though the molecules were transparent. Consequently, the R-T effect has been used to understand sympathetic cooling and to produce cold molecules using natural fermions [10]. Indeed, the appearance of the bound states of the created molecular anions during the electron-complex atom collision at the ground state R-T minimum of the TCS provides an excellent environment and mechanism for breaking up molecular bonds in new molecules creation as well as in anionic catalysis [11]. Contrary to the usual very short-lived temporary anions formation observed in low-energy (0-15 eV) electron collisions with polyatomic molecules, electron-fullerenes interactions are characterized by very long-lived metastable anions formation [12-15].

In electron interactions with fullerene-like molecular systems, simple model potentials are widely used to describe the $C_{60}$ and other fullerene shells [16]. Over the years, beginning with the Dirac bubble potential model representation of the fullerene $C_{60}$ [17, 18], determined by two experimental parameters, viz. its radius and the EA, considerable theoretical investigations have been devoted to understanding the cage environment effect on the confined atoms, including endohedral fullerenes, $A@C_{60}$ [19, 20] and references therein. Most popular among these potentials are the Dirac delta-potential and the attractive short-range spherical shell potential [19, 20]. More recently, a Gaussian-type potential [21] showing smoother shell boundaries and a power-exponential potential [22] with flexible shell boundaries have been introduced. All these potentials have been employed in the photo-detachment of $C_{60}^-$ ions, in photoionization of endohedral atoms, elastic scattering cross sections, charge-particle impact ionization [17,18; 22-37] etc. Other procedures have also been employed to determine the $\Delta$ and $U_0$ parameters of the square potential well [31-34]   At the heart of the generation of these model potentials is the measured EA of $C_{60}$ or the detachment energy of the $C_{60}^-$ negative ion [17, 18]. This is a clear demonstration of the crucial importance of the availability of reliable EAs. Phenomenological potentials for other fullerenes have also been generated similarly [35-37].

The prediction of the distortion of the Xe 4d giant resonance in the photoionization of the Xe@$C_{60}$[17, 26] and the experimental verification of confinement resonances in the photoionization cross section of the 4d subshell of Xe atom in molecular Xe@$C_{60}$ [38] stimulated significant theoretical investigations. Some used the rectangular potential as a fitting potential [39-42] with the parameters selected to best describe the experimental data [38, 42]. However, Amusia and Chernysheva [43] have cautioned against the use of such a procedure. The most attractive and interesting problem in photoionization of Xe@$C_{60}$ is understanding the Xe 4d confinement resonances [17, 18, 22-30, 35-37].

To our knowledge theoretical and experimental low-energy < 10 eV electron elastic scattering TCSs for fullerenes, beyond $C_{60}$ are almost non-existent. Consequently, most of the electron elastic TCSs for the various fullerenes presented here are the first and only. For $C_{60}$ low-energy electron elastic scattering calculations were carried out by Winstead and McKoy [44] within the static-exchange approximation. The Schwinger multichannel method was used for the calculations. Most important in the calculation [44] is the neglect of the polarization and correlation effects. Worth mentioning are the high-level computational calculations, with approximate accounting of polarization, of low-energy elastic cross sections [45-47]. The same group subsequently investigated low-energy resonant structures in electron scattering from the C-20 fullerene [48]. Geometry effects in elastic scattering and capture of electrons by $C_{60}$ have been studied [49]. More recently, Dolmatov *et al* [50, 51] studied electron elastic scattering from endohedral $C_{60}$ while Amusia and Chernysheva [52] investigated scattering phases in electron collision with multi-atomic systems. These recent theoretical studies of the e-$C_{60}$ and the e-A@$C_{60}$ interactions have employed mostly simplified model potentials to represent the $C_{60}$ cage. Experimentally, only low-energy electron elastic scattering DCSs are available, measured by Tanaka *et al* [53]. Interestingly and importantly, the experiment [54] investigated electron impact on gas phase fullerenes $C_{76}$ and $C_{78}$ and obtained the lifetimes of their negative ions in the energy range 10-15 eV(outside the energy range of this investigation). Previously, Elhamidi *et al* [12] investigated low-energy electron attachment in the gas phase $C_{60}$ and $C_{70}$ and identified several resonant states as well as determined the lifetimes of the formed negative ions.

A challenging and lingering problem facing theoretical studies of the interaction of electrons and photons with $C_{60}$ and other fullerenes is the representation of the fullerene cage potential. Recently, Baltenkov *et al* [55] investigated various popular $C_{60}$ shell model potentials and concluded that they generated non-physical charge distributions. Furthermore, these authors [55] noted that the representation of the fullerene shell potential as a model function U(r) is a mere idealization of the real potential of the $C_{60}$ shell. Consequently, the model potentials should not be expected to describe correctly most of the essential features of $C_{60}$ or A@$C_{60}$. This notwithstanding, the square-well type potentials of $C_{60}$ should still be usable as fitting potentials. Very recently, Schrange-Kashenock [56] has modeled the fullerene cage by *ab initio* spherical jellium shell, thereby accounting for the real carbon electron density distribution.

It must be emphasized here that the general subject of electron and photon interactions with clusters, including fullerenes is immense and has experienced considerable investigations, providing a wealth of knowledge to the field, see for example [57-64] and references therein. Our investigation of low-energy electron elastic scattering from selected fullerenes is very focused and provides a new approach to the field populated by many varied expertise, including theoretical methods probing many facets of the underlying physics, thereby providing a rich insight into the field. Therefore, our paper, completely new in the field of clusters/fullerenes, benefitted immensely from the materials in [57-64]. It should be considered as an investigation toward a fundamental understanding of low-energy electron collisions with fullerenes leading to fullerene negative ions formation as resonances. Our approach is also capable of calculating low-lying excited states of fullerene anions; this will be a subject of our next paper. The structure of the paper is as follows: Section 2 presents the brief theory. Sections 3 and 4 give the results and conclusions, respectively.

## 2. Method of calculation

### 2.1 Elastic Cross Section

The Regge-pole (also known as the CAM) method is appropriate for investigating low-energy electron scattering from the fullerenes, resulting in fullerene negative ion formation as resonances since Regge poles, singularities of the S-matrix, rigorously define resonances [65,66]. Embedded in the Regge pole method are the crucial electron-electron correlations and the vital core polarization interaction. These effects are mostly responsible for the existence and stability of typical negative ions. The fundamental quantities which appear in the CAM theories are the energy-dependent positions and residues of Regge poles. Plotting Im L(E) versus Re L(E) (L is the CAM) the well-known and revealing Regge trajectories can be investigated [67]; they probe electron attachment at the fundamental level near threshold since they penetrate the atomic core. Their importance in low-energy electron scattering has been demonstrated in for example [68].

Within the CAM representation of scattering, the Mulholland formula [69] for the electron elastic TCS, wherein is embedded fully the electron-electron correlations, takes the form [70, 71] (atomic units are used throughout):

$$\sigma_{tot} = 4\pi k^{-2} \int_0^\infty Re\,[1 - S(\lambda)]\lambda d\lambda - 8\pi^2 k^{-2} \sum_n Im\,\frac{\lambda_n \rho_n}{1+\exp(-2\pi i \lambda_n)} + I(E) \qquad (1)$$

where S is the S-matrix, k = √(2mE), with m being the mass, $\rho_n$ the residue of the S-matrix at the nth pole, $\lambda_n$ and I(E) contains the contributions from the integrals along the imaginary λ-axis; its contribution is negligible [67]. Here we consider the case for which Im $\lambda_n$<<1 so that for constructive addition, $Re\,\lambda_n \approx 1/2, 3/2, 5/2$ …, yielding $l = Re\,L \cong 0, 1, 2$ …..The significance of Eq. (1) is that a resonance is likely to influence the elastic TCS when its Regge pole position is close to a real integer [71].

### 2.2 Potential

Here it is appropriate to place in perspective the development of the robust and powerful potential, now known as the Avdonina-Belov-Felfli (ABF) potential, and used here for the first time ever for the fullerenes. The crucial core-polarization interaction is incorporated into the ABF potential. When used with Eq.(1) the ABF potential has an impressive history of success in calculating and predicting BEs of both weakly and tenuously bound negative ions formed during low-energy electron collisions with simple and complex atoms.

Here we consider the low-energy behavior of the elastic total cross section (TCS) in the complex angular momentum representation of scattering within the simple approximation, *viz.* the

Thomas–Fermi (T-F) theory, see for example [72, 73] and references therein. The exact form of the needed T-F potential q(r) for an atom is determined from the universal T-F function χ(r) [74] (atomic units are used throughout)

$$q(r) = \frac{-Z\chi(r)}{r} \tag{2}$$

where Z is the atomic number of the target atom and the function χ(r) obeys the non-linear T-F differential equation

$$\frac{d^2\chi(x)}{dx^2} = \frac{1}{\sqrt{x}} \chi^{3/2}(x) \tag{3}$$

The unusual boundary conditions are

$$\chi(x) > 0, \quad x > 0$$
$$\chi(0) = 1$$
$$\chi(x) \sim \frac{144}{x^3}, \quad x \to \infty \tag{4}$$

The variable $x$ is given by $x = r/\mu$, with $\mu = 0.8853 a_0 Z^{-1/3}$ and $a_0$ is the Bohr radius. A transcendental equation, Eq. (3) has only a numerical solution to date. A tabulation of the χ(x) function, also known as the universal T-F function, can be found in for example [75]. The Majorana solution of the T-F equation leads to a semi-analytical series solution [76]. In [77] the simple Padé approximant procedure was demonstrated to produce a remarkably good representation for the T-F exact solution.

In Felfli *et al* [78] the exact function χ(x) was replaced by its rational function approximation in the Tietz context [79]. Within the T-F theory, Tietz used his potential [79] to investigate low-energy electron elastic scattering from many atoms and obtained the analytical expressions for both the total cross sections and the scattering length. In their investigation of Regge poles trajectories for non-singular potentials, Felfli *et al* [78] generated from the approximate function χ(x) a potential of the form, now known as the ABF potential,

$$V(r) = \frac{-Z}{r(1+\alpha Z^{1/3} r)(1+\beta Z^{2/3} r^2)} \tag{5}$$

where α and β are variational parameters. Notably, our choice of the ABF potential, Eq. (5) is adequate as long as we limit our investigation to the near-threshold energy regime, where the elastic cross section is less sensitive to short-range interactions and is determined mostly by the polarization tail. Note also that the ABF potential has the appropriate asymptotic behavior, *viz.* ~ $-1/(\alpha\beta r^4)$ and accounts properly for the polarization interaction at low energies. The advantage of the replacement of the exact function χ(x) by its rational function approximation, namely the well-investigated Eq.(5), is that it is a good analytic function that can be continued to the complex plane.

In particular, the structure of the poles of the ABF potential, Eq. (5) was investigated in the complex plane in [80]. The potential was found to consist of five complex turning points. In contrast to the Tietz approximation, where there are two poles, one of order 1 and the other of order 2, all the poles in the ABF potential are of order 1. Also, contrary to the case of singular potentials considered in [81], the anti-Stokes lines topology for the ABF potential is characterized by a more complicated structure, thereby making the anti-Stokes lines topology very difficult to implement [80].

Over the years the structure and topology in the complex plane of the ABF potential has been investigated extensively in the context of Regge poles trajectories, Re L(E) versus Im L(E) [78, 80, 82, 83]. In the Felfli *et al* [78] studies of Regge poles trajectories the parameters "α" and *"β"* were kept fixed while in Belov *et al* [80] they were varied. Perhaps, the most important investigation and the attendant revelation of the importance of Regge poles trajectories in low-energy electron scattering using the ABF potential were carried out by Thylwe [68]. For atomic Xe the Dirac Relativistic and non-Relativistic Regge trajectories were contrasted near threshold and found to yield essentially the same Re L(E) when the Im L(E) was still very small, see Fig. 2 of [68]. This is a clear demonstration of the insignificant difference between the Relativistic and non-Relativistic calculations at low scattering energies, corresponding to possible electron attachment, leading to negative ion formation as resonances.

The effective potential

$$V(r) = U(r) + L(L+1)/(2r^2),  \qquad (6)$$

is considered here as a continuous function of the variables $r$ and L. The potential, Eq. (5) has been used successfully with the appropriate values of α and β. When the TCS as a function of "*β*" has a resonance [71] corresponding to the formation of a stable bound negative ion, this resonance is longest lived for a given value of the energy which corresponds to the BE of the system (for ground state collisions). This was found to be the case for all the systems we have investigated thus far and fixes the optimal value of "*β*" for Eq. (5).

For the numerical evaluation of the TCSs and the Mulholland partial cross sections, we solve the Schrödinger equation for complex values of L and real, positive values of E

$$\psi'' + 2\left(E - \frac{L(L+1)}{2r^2} - U(r)\right)\psi = 0, \qquad (7)$$

with the boundary conditions:

$$\psi(0) = 0,$$
$$\psi(r) \sim e^{+i\sqrt{2E}r}, r \to \infty. \qquad (8)$$

We note that Eq. (8) defines a bound state when k ≡ √(2E) is purely imaginary positive. We calculate the S-matrix, S(L, k) poles positions and residues of Eq. (7) following a method similar to that of Burke and Tate [84]. In the method the two linearly independent solutions, $f_L$ and $g_L$, of the Schrödinger equation are evaluated as Bessel functions of complex order and the S-matrix, which is defined by the asymptotic boundary condition of the solution of the Schrödinger equation, is thus evaluated. Further details of the calculation may be found in [84].

In Connor [85] and Ref. [67] the physical interpretation of Im L is given. It corresponds inversely to the angular life of the complex formed during the collision. A small Im L implies that the system orbits many times before decaying, while a large Im L value denotes a short-lived state. For a true bound state, namely E < 0, Im L = 0 and therefore the angular life, 1/[Im L], implying that the system can never decay. Im L is also used to differentiate subtleties between the ground and the excited states of the negative ions formed as resonances during the collisions.

## 3. Results.

Figure 1 presents the variation with E(eV) of low-energy electron elastic scattering TCS for $C_{60}$. Just as the ground state TCS for e - Au scattering [9], insert, it is characterized by two R-T minima, a shape resonance and a dramatically sharp resonance, representing the formation during the collision of a long-lived ground state $C_{60}^-$ negative ion. The first R-T minimum appears at about 0.7 eV, followed by the shape resonance at about 1.0 eV, and the second absolute minimum is located at about 2.5 eV. The sharp resonance at 2.66 eV represents the formed ground state $C_{60}^-$ anion during the collision; this binding energy (BE) of the $C_{60}^-$ anion is identified with the measured EA of $C_{60}$. Indeed, it was pointed out by Johnson and Guet [86] that the appearance of the R-T minima manifests that the polarization interaction has been accounted for adequately. Contrary to the e-Au scattering case, the e-$C_{60}$ TCS also exhibits a shape resonance near threshold at around 0.065 eV. The shape resonance at 1.0 eV and the second R-T minimum at about 2.5 eV in the TCS for $C_{60}$ were also identified in the calculated partial cross sections of Winstead and McKoy [44]. We emphasize here that there are no external experimental or other theoretical inputs in our calculation of the TCS. Only the *"β"* parameter of the polarization potential is optimized. When the dramatically sharp resonance appears as in the Fig. 1, it is indicative of electron attachment. In this case, like in many previously investigated simple and complex atoms, the BE corresponds to the measured EAs of $C_{60}$ [87-89, 99]; the agreement between the measured EAs and our BE is outstanding.

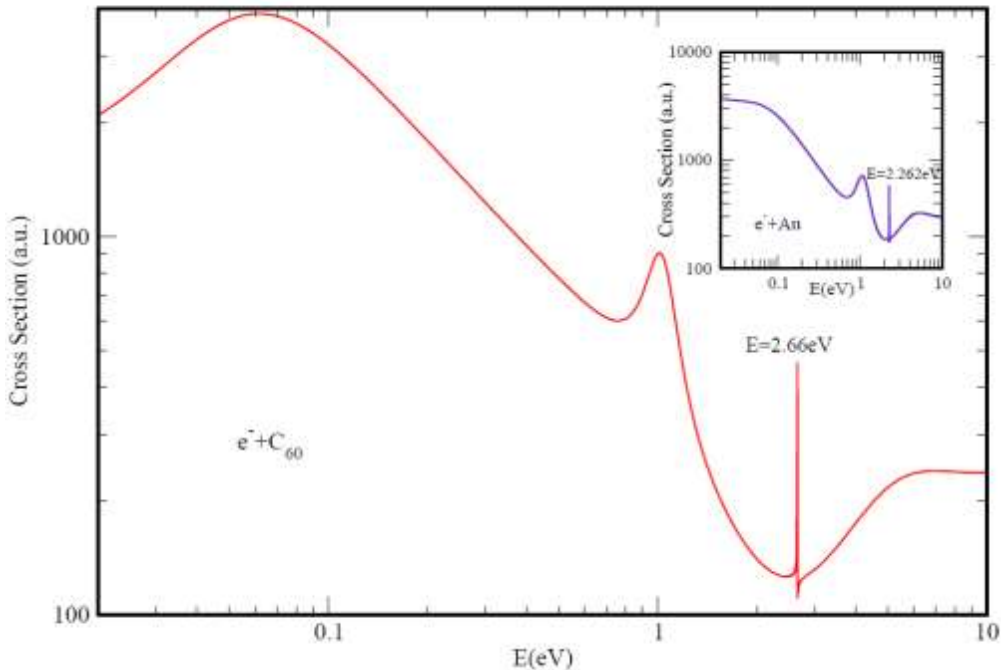

**Figure 1:** Variation with E(eV) of low-energy electron elastic scattering TCS for $C_{60}$. The insert corresponds to the TCS for e-Au scattering for comparison.

Figures 2(a), (b), (c) and (d) demonstrate the variation with E(eV) of the calculated TCSs for the e- $C_{70}$, e- $C_{76}$, e- $C_{78}$, and e- $C_{84}$ scattering, respectively. Clearly, as in the case of the e- $C_{60}$ scattering, all the TCSs are characterized by SRs, R-T minima and the dramatically sharp resonances appearing at the second R-T minima of their TCSs. Their values are summarized in Table 1. Table 2 compares our calculated BEs of the resultant fullerene negative ions formed during the collisions with the measured

EAs for $C_{70}$, $C_{76}$, $C_{78}$ and $C_{84}$ [88, 90, 91, 92]. Indeed, as with the case of the e- $C_{60}$ scattering the agreement is outstanding. It is noted that in the reference [93] additional to the measurement of the EA of $C_{84}$, the important EAs of Gd@$C_n$ (n=74, 76, 78, 80, 82) were measured as well.

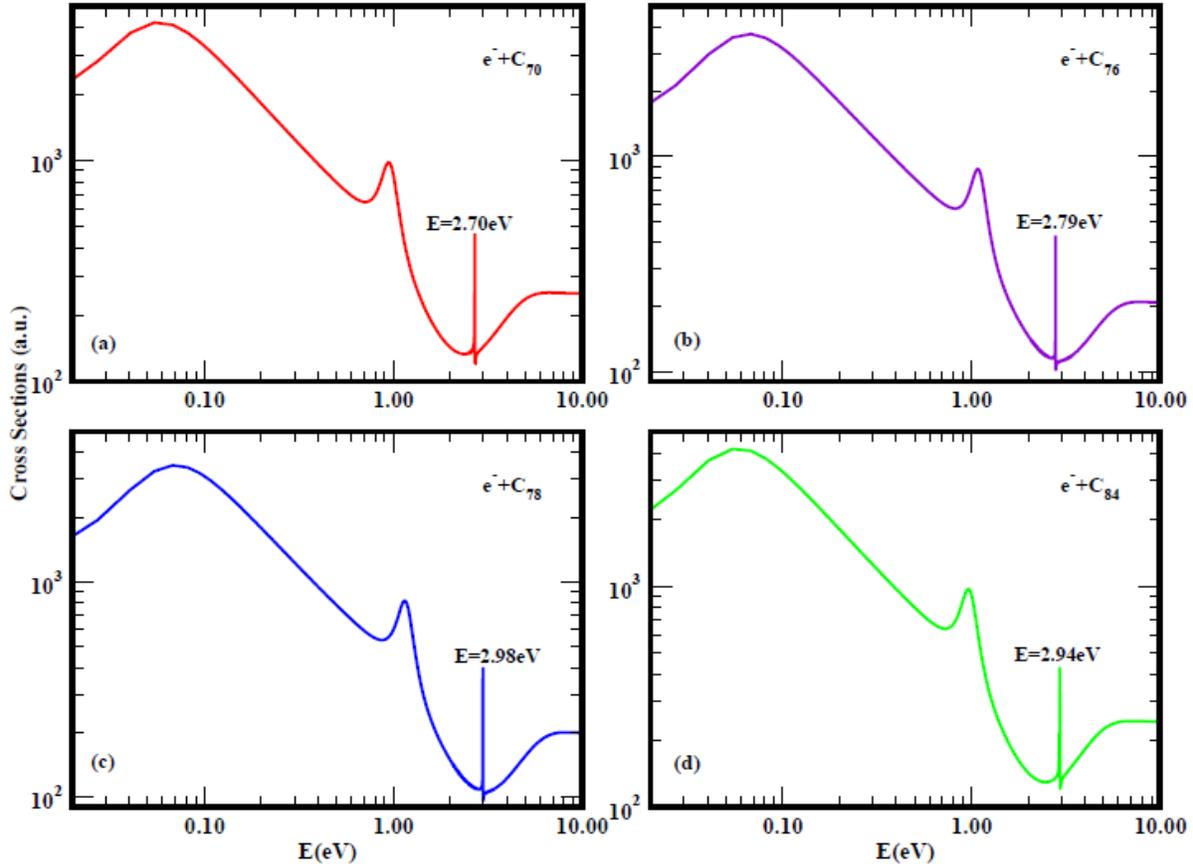

**Figures 2**(a), (b), (c) and (d): Variation with E(eV) of the calculated TCSs for the e- $C_{70}$, e- $C_{76}$, e- $C_{78}$, and e- $C_{84}$ scattering, respectively.

In Figs. 3(a), (b), (c) and (d) is presented the TCSs for the electron scattering from $C_{86}$, $C_{90}$, $C_{92}$ and $C_{180}$ fullerenes, respectively. The fundamental characteristic structure of the e- $C_{60}$ scattering TCS clearly permeates these TCSs. The values of their SRs, R-T minima and the BEs of the resultant negative ions are summarized in Table 1 as well. From Table 2 it is noted that the calculated BEs for the $C_{86}$, $C_{90}$ and $C_{92}$ are very close to those estimated by the experiment to be $\geq$ 3.0 eV[90]. For the $C_{180}$ fullerene molecule, the calculated BE of its resultant negative ion is 2.54 eV which agrees excellently with the value of 2.611 eV calculated by Lei Xu *et al* [94]. For these fullerenes there are no electron elastic scattering TCSs available to our knowledge to compare with.

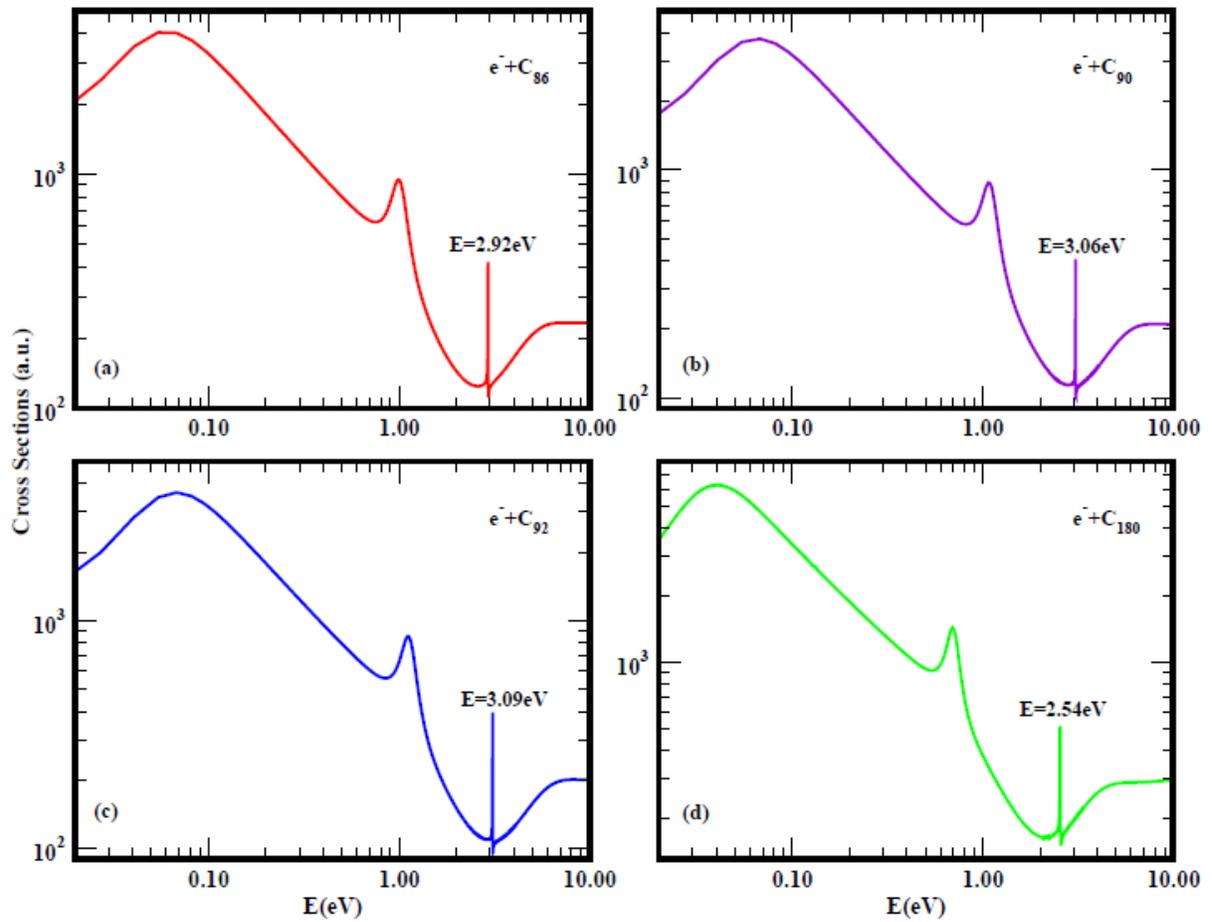

**Figures 3**(a), (b), (c) and (d): Variation with E(eV) of the calculated TCSs for the $C_{86}$, $C_{90}$, $C_{92}$ and $C_{180}$ fullerenes, respectively

In [90] the EAs of the $C_{82}$, $C_{74}$ and $C_{80}$ fullerene molecules were measured. In Figs. 4(a), (b), (c) and (d) is presented the variation with E(eV) of the calculated electron elastic scattering TCSs for $C_{82}$, $C_{74}$, $C_{80}$, and $C_{54}$, respectively. The selection of the $C_{54}$ fullerene is dictated to by the interesting finding that the quantum confinement resonances are destroyed if the shape of the fullerene deviated significantly from a sphere [37]. All the TCSs in Fig. 4 exhibit the fundamental structure of the e- $C_{60}$ scattering TCS. Consequently, it can be safely conjectured that many fullerenes will exhibit BEs of their negative ions close to those measured in [87-93] and [98, 99]. The extracted data for these fullerenes are also presented in Table 1 and in Table 2 they are compared with the measured EA values. Comparing the BEs of the fullerene negative ions in Tables 1 and 2, one sees that the BE of the $C_{54}^-$ fullerene negative ion is larger than that of the $C_{60}^-$ and $C_{240}^-$. The results are consistent with the finding that the $C_{240}$ fullerene is closest to a sphere compared to $C_{60}$ and, by comparison, $C_{54}$ deviates the most from a sphere [37].

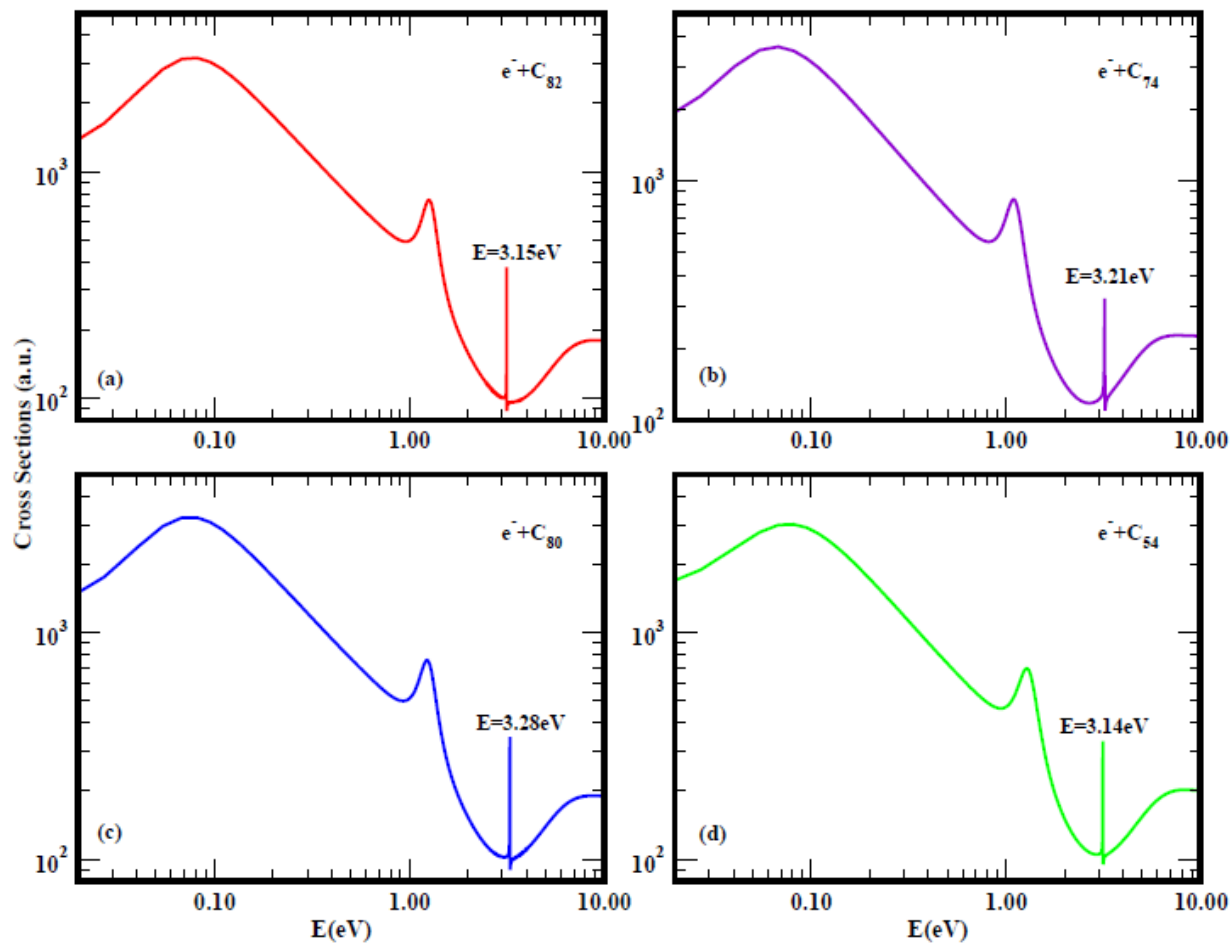

**Figures 4**(a), (b), (c) and (d): Variation with E(eV) of the calculated electron scattering TCSs for the $C_{82}$, $C_{74}$, $C_{80}$, and $C_{54}$ fullerenes, respectively

Although not plotted in the paper, the relevant data for the e- $C_{58}$ and e- $C_{240}$ scattering are also included in Tables 1 and 2. Our extracted BE for the $C_{240}^{-}$ negative ion agrees excellently with the calculated EA [95] but disagrees significantly with the calculated value of [97]. Also, our calculated BE for the $C_{58}^{-}$ negative ion is 2.41 eV. Interestingly, this value is smaller than that of $C_{60}^{-}$ and, as expected, than that of $C_{54}^{-}$. To our knowledge there are no measured or calculated EAs available to compare our values with. Indeed, measurements of the EAs of $C_{54}$ and $C_{58}$ would help to verify our values for these molecules. Suffice to state that these BEs for the $C_{54}^{-}$ and $C_{58}^{-}$ fullerene negative ions should be useful in the generation of the standard model potentials needed in the investigations of electron and photon interactions with A@$C_{54}$ and A@$C_{58}$ endohedral fullerene molecules.

Before we conclude the paper it is appropriate to remark on the general structures of the fullerene TCSs displayed in the various figures in the context of fullerene negative ion catalysis. This is motivated by their resemblance to the TCS of atomic Au, viz. the appearance of the BEs of the resultant fullerene negative ions at the second R-T minima of the TCSs. The experiments of Hutchings and collaborators [100-102] previously used supported Au, Pd and Au-Pd nanoparticle catalysts for the direct synthesis of hydrogen peroxide from $H_2$ and $O_2$ [100, 101]. The experiments also found that the addition of Pd to the Au catalyst increased the rate of $H_2O_2$ synthesis significantly as well as the concentration of the $H_2O_2$ formed. More recently, Freakley *et al*. [102] reported a significant improvement in the direct synthesis of $H_2O_2$ when using a Pd-Sn alloy. The results of the experiments [100, 101] were explained through the formation of the anionic molecular complexes $Au^-(H_2O)_{1,2}$ and $Pd^-(H_2O)_{1,2}$, resulting in the disruption of the H-O bonds [11]. The large electron affinity of Au played an essential role as well. It would be interesting and important to replace the two Au and Pd nanocatalyst with a single catalyst in the experiments [100, 101].

In [103] the importance of the fullerene EAs in catalysis of organic solar cell materials was investigated and confirmed. Consequently, here we explored the possibility of identifying the mechanism of fullerene negative ion catalysis hoping to find a long-lived fullerene metastable negative ion resonance near the ground state resonance. So, we calculated the TCS for the first excited state using $C_{92}$ as an example. Figure 5 shows the $C_{92}$ TCS and that of the first excited state of $C_{92}$, insert. Indeed, the metastable $C_{92}^-$ resonance appears within the second R-T minimum of the ground state TCS of the $C_{92}$ fullerene; the two resonances are separated by about 0.4 eV and will contribute to each other's catalytic performance. This is the first identification of a single system with multiply functionalized catalyzing capability. We expect that many of the fullerene molecules TCSs will exhibit similar structures as that of the $C_{92}$ fullerene. We recommend both experimental and theoretical investigations of the resonances in the TCSs of fullerenes in the energy region of the second R-T minima.

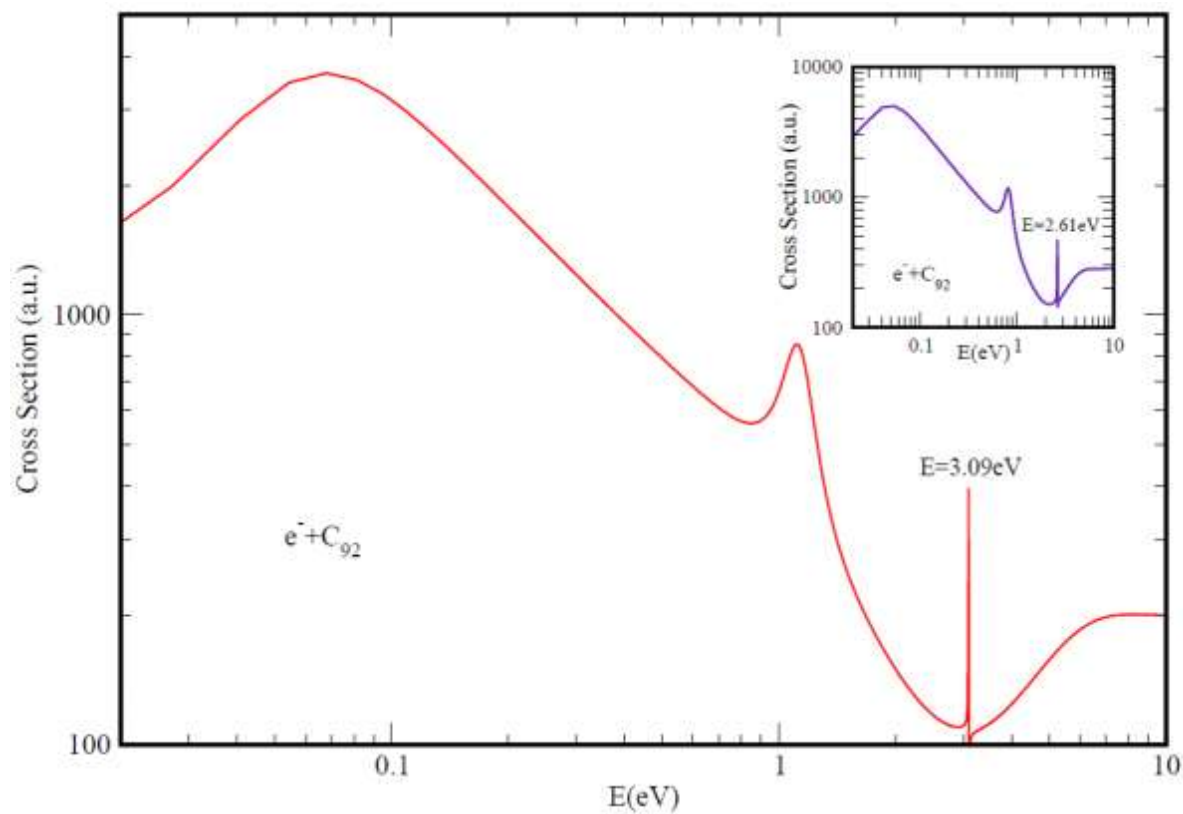

**Figure 5:** Variation with E (eV) of low-energy electron elastic scattering TCS for $C_{92}$ fullerene. The insert corresponds to the TCS for the excited $C_{92}$ fullerene whose negative ion BE is 2.61eV.

**Table 1.** Ramsauer–Townsend (R-T) minima and Shape Resonances (SRs) in the elctron scattering cross sections for the fullerenes $C_{54}$ through $C_{240}$ as well as the Binding Energies (BEs) of their resultant negative ions and the optimized 'β' parameters of the polarization potential. RT-1 and RT-2 represent the first and second and R-T minimum, respectively.

| System | β | BE (eV) | SR-1 (eV) | RT-1 (eV) | SR-2 (eV) | RT-2 (eV) |
|---|---|---|---|---|---|---|
| $C_{54}$ | 0.0567 | 3.14 | 0.080 | 0.94 | 1.28 | 2.92 |
| $C_{58}$ | 0.0447 | 2.49 | 0.054 | 0.71 | 0.94 | 2.30 |
| $C_{60}$ | 0.0479 | 2.66 | 0.061 | 0.76 | 1.02 | 2.46 |
| $C_{70}$ | 0.04576 | 2.70 | 0.054 | 0.71 | 0.95 | 2.38 |
| $C_{74}$ | 0.0513 | 3.21 | 0.068 | 0.82 | 1.10 | 2.68 |
| $C_{76}$ | 0.0539 | 2.79 | 0.082 | 0.82 | 1.09 | 2.68 |
| $C_{78}$ | 0.0568 | 2.98 | 0.068 | 0.87 | 1.16 | 2.85 |
| $C_{80}$ | 0.0598 | 3.28 | 0.081 | 0.92 | 1.24 | 3.07 |
| $C_{82}$ | 0.0627 | 3.15 | 0.081 | 0.95 | 1.26 | 3.06 |
| $C_{84}$ | 0.04725 | 2.94 | 0.054 | 0.72 | 0.97 | 2.47 |
| $C_{86}$ | 0.0495 | 2.92 | 0.054 | 0.75 | 1.00 | 2.58 |
| $C_{90}$ | 0.0542 | 3.06 | 0.068 | 0.82 | 1.07 | 2.81 |
| $C_{92}$ | 0.0566 | 3.09 | 0.068 | 0.84 | 1.11 | 2.91 |
| $C_{180}$ | 0.0401 | 2.54 | 0.040 | 0.54 | 0.69 | 2.13 |
| $C_{240}$ | 0.0445 | 2.41 | 0.041 | 0.54 | 0.71 | 2.32 |

**Table 2.** Comparison between the present BEs (eV) and the measured and other theoretically calculated EAs for the investigated fullerenes, $C_{54}$ through $C_{240}$.

| System | EA-Expt. [90] (eV) | EA-Other Expts. (eV) | BE-Present (eV) | EA-Other Theory (eV) |
|---|---|---|---|---|
| $C_{60}$ | | 2.666±0.001[88]; 2.683(0.008) [99]; 2.65[87]; 2.6835±0.0006 [87]; 2.69[89] | 2.66 | 2.57[96] 2.23[95] |
| $C_{70}$ | 2.72 ± 0.05 | 2.676±0.001 [88] 2.765(0.01)[98] | 2.70 | |
| $C_{74}$ | 3.28 ± 0.07 | 3.28±0.07 [92] | 3.21 | |
| $C_{76}$ | 2.88 ± 0.05 | 2.89±0.05[92]; 2.975±0.01[91] | 2.79 | |
| $C_{78}$ | 3.01 ± 0.07 | 3.10±0.06[92]; 3.10±0.01[91] | 2.98 | |
| $C_{80}$ | 3.17 ± 0.06 | 3.17±0.06 [92] | 3.28 | |
| $C_{82}$ | 3.14 ± 0.06 | 3.14±0.06[92] | 3.15 | 3.37[96] |
| $C_{84}$ | 3.05 ± 0.08 | 3.14(6)[92]; 3.185±0.01[91] 3.16[93] | 2.94 | |
| $C_{86}$ | ≥ 3.0 | | 2.92 | |
| $C_{90}$ | ≥ 3.0 | | 3.06 | |
| $C_{92}$ | ≥ 3.0 | | 3.09 | |
| $C_{180}$ | | | 2.54 | 2.611[94] |
| $C_{240}$ | | | 2.41 | 2.32[95] 3.81[97] |
| $C_{54}$ | | | 3.14 | |
| $C_{58}$ | | | 2.49 | |

## 4. Summary and Conclusions

The Regge pole method has been used for the first time together with the robust ABF potential to investigate the electron elastic TCSs of $C_{54}$ through $C_{240}$ fullerenes in the electron impact energy range $0.02 \leq E \leq 10.0$ eV. The TCSs are found to be characterized generally by R-T minima, shape resonances and dramatically sharp resonances representing long-lived ground state fullerene negative ion formation. From the calculated TCSs we extracted the BEs of the negative ions formed during the electron–fullerene collision as Regge resonances. The agreement between our calculated BEs and the measured EAs is generally outstanding. For the case of the $C_{92}$ fullerene we also calculated the TCS for the first excited state and found it to be a long-lived excited (metastable) state. The location side by side of the BEs of the ground and the metastable state fullerene anions at the second R-T minimum of the ground state TCS of $C_{92}$ (see Fig. 5) makes the $C_{92}$ a powerful multiplecatalyst through its negative ions (we expect similar structures for many other fullerenes). It must be emphasized here that the Regge pole method requires no external input data, either experimental or theoretical to achieve these remarkable results.

Since the EA is a sensitive and an important measured and/or calculated quantity, it provides a stringent test of theoretical calculations when the calculated EAs are compared with those from reliable measurements. Fortunately, for many of the fullerenes considered in this paper high quality measured EAs are available. Looking at Table 2, the experimental EAs are concentrated in the range $C_{60}$ through $C_{84}$ fullerenes; we could not find the EAs for fullerenes in the wider range $C_{86}$ through $C_{240}$. The situation is even worse with the theoretical determination of the EAs of fullerenes in general; to our knowledge theoretical EAs are available only for the fullerenes that are closer to a sphere, namely $C_{60}$, $C_{180}$ and $C_{240}$.

Currently, we are investigating additional resonance structures representing metastable fullerene negative ion formations within the energy region of the second R-T minima of the TCSs of the tabulated fullerenes. Suffice to state that the present BEs of the tabulated fullerenes in Table 1 should help in the construction of the popular model potential wells for those fullerenes and the corresponding endohedral fullerenes. In conclusion, the success of the Regge pole method in calculating the electron elastic TCSs of the investigated fullerenes is due mainly to the fact that the electron-electron correlations are fully embedded in the Mulholland formula while the ABF potential incorporates the polarization interaction.

## Acknowledgments
Research was supported by the US DOE, Division of Chemical Sciences, Geosciences and Biosciences, Office of Basic Energy Sciences, Office of Energy Research and CAU CFNM, NSF-CREST Program. The computing facilities of the National Energy Research Scientific Computing Center are greatly appreciated.